\documentstyle[epsfig,12pt,a4p,subfigure]{article}

\newcommand{\be}{\begin{equation}}
\newcommand{\ee}{\end{equation}}
\newcommand{\bea}{\begin{eqnarray}}
\newcommand{\eea}{\end{eqnarray}}
\newcommand{\bean}{\begin{eqnarray*}}
\newcommand{\eean}{\end{eqnarray*}}

\newcommand{\gapproxeq}{\lower
.7ex\hbox{$\;\stackrel{\textstyle >}{\sim}\;$}}
\newcommand{\lapproxeq}{\lower
.7ex\hbox{$\;\stackrel{\textstyle <}{\sim}\;$}}

\begin{document}
\begin{titlepage}
\begin{tabbing}
wwwwwwwwwwwwwwwright hand corner using tabbing so it looks neat and in \= \kill
\> {27 April 2000}
\end{tabbing}
\baselineskip=18pt
\vskip 0.7in
\begin{center}
{\bf \LARGE The mixing of the $f_0(1370)$, $f_0(1500)$ and
$f_0(1710)$ and the search for the scalar glueball}\\
\vspace*{0.9in}
{\large Frank E. Close}\footnote{\tt{e-mail: F.E.Close@rl.ac.uk}} \\
\vspace{.1in}
{\it Rutherford Appleton Laboratory}\\
{\it Chilton, Didcot, OX11 0QX, England}\\
\vspace{0.1in}
{\large Andrew Kirk}\footnote{\tt{e-mail: ak@hep.ph.bham.ac.uk}} \\
{\it School of Physics and Astronomy}\\
{\it Birmingham University}\\
\end{center}
\begin{abstract}
For the first time a complete data set
of the two-body decays of
the $f_0(1370), f_0(1500)$ and $f_0(1710)$ into all pseudoscalar mesons
is available.
The implications of these data
for the flavour content for these three $f_0$ states is studied.
We find that they are in accord with the hypothesis
that the scalar glueball
of lattice QCD mixes with the $q \overline q$ nonet that also exists
in its immediate vicinity. We show that this
solution also is compatible with the relative production
strengths of the $f_0(1370), f_0(1500)$ and $f_0(1710)$
in $pp$ central production, $p \bar{p}$ annihilations
and $J/\psi$ radiative decays.
\end{abstract}
\end{titlepage}
\setcounter{page}{2}
\par

The best estimate for the masses of glueballs comes from
lattice gauge theory calculations which in the quenched approximation show
\cite{re:lgt}
 that the lightest glueball has $J^{PC}$~=~$0^{++}$ and that
its mass should be in the range
$1.45-1.75$ GeV. While the lattice remains immature for predicting glueball
decays,
Amsler and Close \cite{re:AC} have noted that in lattice
inspired models,
such as the flux tube \cite{re:IP}, glueballs will mix strongly
with nearby $q \overline q$ states with the same
$J^{PC}$~\cite{anisovich}.
This will lead to three isoscalar states of the same
$J^{PC}$ with predictable patterns of
decay branching ratios \cite{re:AC,re:FC89,re:CFL}. Such mixing ideas have been
applied recently to the three states in the
glueball
mass region - the
$f_0(1370)$, $f_0(1500)$ and $f_0(1710)$ \cite{re:AC,re:CFL,Wein,mix,Genov}.

\par

While these papers at first sight differ in detail, nonetheless their
conclusions share some common robust features. The flavour content of
the states have the $n \overline n$ and $s \overline s$ in phase
(SU(3) singlet tendency) for the $f_0(1370)$ and $f_0(1710)$, and
out of phase (octet tendency) for the  $f_0(1500)$ (that such a pattern is
natural is discussed in ref. \cite{FC00}). In general
these mixings will negate the naive
folklore  that glueball decay branching ratios
would be independent of the quark composition
of the final state mesons after taking into account phase space effects,
so called ``flavour blind decays".

\par
Recently ref.\cite{mix} has used some of the observed branching ratios
as input to constrain the mixing pattern. The results here too agree
with the generic structure found in refs. \cite{re:AC,re:CFL,Wein,FC00}.
Most recently, and subsequent to the above work,
the WA102 collaboration has now published~\cite{etaetapap},
for the first time in a single experiment, a complete data set
for the decay branching
ratios of the
$f_0(1370)$, $f_0(1500)$ and
$f_0(1710)$ to all pseudoscalar meson pairs.
These data will be our point of departure.
Using methods similar to those proposed in ref.~\cite{mix}
we investigate the implications of the WA102 data for
the glueball-quarkonia content of the $f_0(1370)$, $f_0(1500)$ and
$f_0(1710)$. This moves the debate forwards in the following ways:

\begin{itemize}
\item
It highlights the sensitivity of the mixings to the input data.
\item
It exposes some assumptions, both explicit and implicit,
in the analysis of ref. \cite{mix} that can be improved upon.
\item
It allows for the direct decay of glueballs into $\eta$ and $\eta '$,
which were not manifested in leading order in ref.\cite{mix}
even though there are reasons to suspect that they could be
important \cite{re:AC,gershtein}.
\end{itemize}

\par
In order to unfold the production
kinematics we use the invariant decay couplings~($\gamma_{ij}$)
for the observed decays, namely
we express the partial width~($\Gamma_{ij}$) as~\cite{re:AC}
\begin{equation}
\Gamma_{ij}=\gamma^2_{ij}|F_{ij}(\vec{q})|^2S_p(\vec{q})
\label{eq:a}
\end{equation}
where $S_p(\vec{q})$ denotes the phase space and $F_{ij}(\vec{q})$
are form factors appropriate to exclusive two body decays.
 Here we have followed ref.~\cite{re:AC}
and have chosen
\begin{equation}
|F_{ij}(\vec{q})|^2 = q^{2l}exp(-q^2/8\beta^2)
\label{eq:b}
\end{equation}
where $l$ is the angular momentum of the final state with daughter momenta
$q$
and we have used $\beta$~=~0.5~GeV/c~\cite{re:AC}.
The $f_0(1500)$ lies very near to threshold in the $\eta \eta^\prime$
decay mode, therefore we have used an average value of $q$ (190~MeV/c)
derived from a fit to the $\eta \eta^\prime$ mass spectrum.
\par
The branching ratios measured by the WA102 experiment for
the $f_0(1370)$, $f_0(1500)$ and $f_0(1710)$ are given in
table~\ref{ta:BR}.
The invariant decay couplings ($\gamma_{ij})$ are related to the
relevant decay amplitudes $M_{ij}$ by
\begin{equation}
\gamma^2_{ij} = c_{ij}|M_{ij}|^2
\label{eq:c}
\end{equation}
where $c_{ij}$ is a weighting factor arising from the sum over the various
charge
combinations, namely 4 for $K \overline K$, 3 for $\pi \pi$, 2 for
$\eta \eta^\prime$ and 1 for $\eta \eta$ for isoscalar decays.
If in the decay of some state the ratios of the decay amplitudes squared
($|M_{ij}|^2$) for the $\eta \eta$/$\pi \pi$ and $\eta \eta/K \overline K$
are simultaneously greater than unity, then this state cannot
be a quarkonium decay (see fig.~3 of ref.~\cite{re:AC}
and also ref.~\cite{re:FC89}).
Table~\ref{ta:amp} gives these ratios for
the $f_0(1370)$, $f_0(1500)$ and $f_0(1710)$, as abstracted using
eqs.(\ref{eq:a}), (\ref{eq:b}) and (\ref{eq:c}). As can be seen the ratios
for the $f_0(1710)$ argue either for non-$q \overline q$
content in this meson or for some further dynamical suppression of
the $\pi \pi$ mode, say, as may occur for special values of parameters
in some specific models \cite{nodes}.
\par
In the $|G\rangle=|gg\rangle$, $|S\rangle=|s\bar{s}\rangle$,
$|N\rangle=|u\bar{u}+d\bar{d}\rangle/\sqrt{2}$ basis,
the mass matrix describing the mixing of a glueball and
quarkonia can be written as follows \cite{Wein}:
\begin{equation}
M=\left( \begin{array}{ccc}
M_G & f &
\sqrt{2}f\\
f & M_S & 0\\
\sqrt{2}f & 0 & M_N
\end{array}\right),
\label{eq:d}
\end{equation}
where $f=\langle G|M|S\rangle=\langle G|M|N\rangle/\sqrt{2}$ represents the
flavour independent mixing strength between the glueball and quarkonia
states. $M_G$, $M_S$ and $M_N$ represent the masses of the bare states
$|G\rangle$, $|S\rangle$ and $|N\rangle$, respectively.
Following refs.~\cite{re:AC,Wein,mix} we assume the mixing
is strongest between the glueball and nearest $q \overline q$
neighbours. With the lattice (in the quenched approximation)
predicting the glueball mass to be in the $1.45-1.75$~GeV region, this
naturally led these papers to focus on
 the physical states $|f_0(1710)\rangle$,
$|f_0(1500)\rangle$ and
$|f_0(1370)\rangle$ as the eigenstates of $M$ with the eigenvalues of
$M_{1}$, $M_{2}$ and $M_{3}$, respectively. (An alternative picture could
involve the states $f_0(1500), f_0(1710), f_0(2000)$;
we do not discuss this in the present
paper).
The three physical states can be read as
\begin{equation} \left( \begin{array}{ccc}
|f_0(1710)\rangle\\
|f_0(1500)\rangle\\
|f_0(1370)\rangle
\end{array}\right)
=U\left(\begin{array}{ccc}
|G\rangle\\
|S\rangle\\
|N\rangle
\end{array}\right)
=\left(\begin{array}{ccc}
x_1 & y_1& z_1\\
x_2& y_2& z_2\\
x_3 & y_3 & z_3
\end{array}\right)
\left(\begin{array}{ccc}
|G\rangle\\
|S\rangle\\
|N\rangle
\end{array}\right),
\label{eq:e}
\end{equation}
where
\begin{equation}
U=\left(\begin{array}{ccc}
(M_1-M_S)(M_1-M_N)C_1&
(M_1-M_N)fC_1& \sqrt{2}(M_1-M_S)fC_1\\
(M_2-M_S)(M_2-M_N)C_2&
(M_2-M_N)fC_2& \sqrt{2}(M_2-M_S)fC_2\\
(M_3-M_S)(M_3-M_N)C_3&
(M_3-M_N)fC_3& \sqrt{2}(M_3-M_S)fC_3
\end{array}\right)
\label{eq:f}
\end{equation}
with $C_{i(i=1,~2,~3)}=[(M_i-M_{S})^2(M_i-M_N)^2+
(M_i-M_N)^2f^2+2(M_i-M_S)^2f^2]^{-\frac{1}{2}}$.

Ref.\cite{mix} considered the following three
hadronic decay paths for the
$f_0(1370)$, $f_0(1500)$ and $f_0(1710)$:

(i) the direct coupling of the quarkonia
component of the three states
to the final
pseudoscalar mesons ($PP$) (fig.~\ref{fi:1}a),

(ii) the coupling
through two intermediate gluons, $n \overline n
(s \overline s) \to gg \to s \overline s (n \overline n) \to PP$
(fig.~\ref{fi:1}b), with
$r_1$ representing the ratio
of the effective coupling strength of this mode to that of the
mode (i);

(iii) the flavour independent coupling of the
glueball component $ gg \to q_i \overline {q_i}$ with subsequent
decay $q_i \overline {q_i} \to PP$ (fig.~\ref{fi:1}c); with
$r_2$ representing the ratio of this mode to (i)

We propose that this is inconsistent. Specifically, the modes (ii)
and (iii) as described above are what have already been accounted for
in generating the mixed states and so $r_1$ and $r_2$ should be set
to zero. This may be seen by comparing the definitions of eq.(\ref{eq:f})
with the corresponding expressions in perturbation theory
(as e.g. eqs.(27-31) in ref \cite{re:AC} extended to second order).
For example, eq.(27) of
ref. \cite{re:AC} (where $C_G$ denotes the normalisation factor)

\begin{equation}
(C_G)^{-1} f_0(G) =
|G\rangle +
          \frac{|s \bar{s}\rangle \langle s \bar{s} | M | G \rangle}
                       {M_{G} - M_{s \bar{s}}}
+ \frac{|n \bar{n}\rangle \langle n \bar{n} | M | G \rangle}
                       {M_{G} - M_{n \bar{n}}}
\label{eq:g}
\end{equation}
may be written in the form of eq.(6), with $M_G \equiv M_1$, and
$N \equiv n \bar{n}$, $S \equiv {s \bar{s}}$

\begin{eqnarray}
(M_G - M_{n \bar{n}}) (M_G - M_{s \bar{s}}) f_0(G) &=
&(M_1 - M_N) (M_1 - M_S)C_1|G\rangle +
          (M_1 - M_N) f C_1|s \bar{s}\rangle \nonumber \\
&&+ (M_1 - M_S) \sqrt{2} f C_1 |n \bar{n}\rangle
\label{eq:h}
\end{eqnarray}
This shows how the coefficients $y_1,z_1$ are equivalent
to the perturbation
which is in turn driven by fig.~\ref{fi:1}c).
Thus it is double counting to invoke
this same figure with strength $``r_2"$ to describe $gg$ decays via
$q \overline q$ intermediate states, having already used it to
compute the mixing of those same $q \overline q$ in the Fock state.
Similar remarks apply to the $s \overline s \to n \overline n$ mixing
in second order perturbation theory and fig.~\ref{fi:1}b defined as
$r_1$.

However, there are additional pathways that have not been
allowed for in ref.~\cite{mix}.
First there is the role of $gg \to qq \overline {qq}$
as in fig.~\ref{fi:1}d. These may have important resonant
contributions in the region below 1 GeV but are expected to be
primarily a continuum in the 1.5~GeV region of interest here~\cite{Jaf00};
we shall
approximate them by assuming flavour independent couplings. (In a more
sophisticated analysis one could incorporate threshold effects in
the $PP$ channels that overlap the $qq \overline {qq}$
configurations~\cite{tornpenn};
we do not discuss this further in this first look).
The resulting amplitudes can be obtained from eqs.(A4) of ref.~\cite{re:AC}
and have the same structure as those of (iii) above. Hence a non-zero
$r_2$ is restored, though its interpretation differs from ref \cite{mix}.

Finally, following ref \cite{re:AC,gershtein}, we allow for

 (iv) the direct coupling of the glue in the initial state to
isoscalar mesons (i.e. $\eta \eta$ and $\eta \eta^\prime$ decays).
As in ref \cite{re:AC}, we assume chiral symmetry such that the coupling
to the $s \overline s$ content of the $\eta,\eta'$ dominates and allow
$r_3$ to be the ratio
of mode (iv) to (i).


\par

The three decay diagrams considered are shown in fig.~\ref{fi:2}a-c.
Performing an elementary SU(3) calculation gives the
reduced partial widths in table~\ref{ta:RPW},
where $\alpha=(\cos\phi-\sqrt{2}\sin\phi)/\sqrt{6}$,
$\beta=(\sin\phi+\sqrt{2}\cos\phi)/\sqrt{6}$, and
$\phi$ is the mixing angle of
$\eta$ and $\eta\prime$.
The relevant
expressions follow from appendix A in ref. \cite{re:AC} with
$\rho = R =1$ in the case of flavour independence of the direct
couplings.
The predicted branching ratios can then be calculated using
eqs.(\ref{eq:a}) and (\ref{eq:b}).
\par
We then perform a $\chi^2$ fit based on the branching ratios
given in table~\ref{ta:BR},
where we have required that the matrix $U$ in eq.(\ref{eq:e}) is
unitary, which applies an additional 6 constraints to the fit.
As input we use the masses of the
$f_0(1500)$ and $f_0(1710)$.
In this way eight parameters, $M_G$,
$M_N$, $M_S$, $M_3$, $f$, $r_2$, $r_3$ and $\phi$
are determined from the fit.
The parameters determined from the solution
with the lowest $\chi^2$
are presented in table~\ref{ta:res1} and the fitted branching ratios
together with the $\chi^2$ contributions of each are given in
table~\ref{ta:BR}.
\par

The physical states $|f_0(1710)\rangle$, $|f_0(1500)\rangle$ and
$|f_0(1370)\rangle$ are found to be
\begin{equation}
|f_0(1710)\rangle=0.39|G\rangle+0.91|S\rangle+0.14|N\rangle,
\label{eq:i}
\end{equation}
\begin{equation}
|f_0(1500)\rangle=-0.69|G\rangle+0.37|S\rangle-0.62|N\rangle,
\label{eq:j}
\end{equation}
\begin{equation}
|f_0(1370)\rangle=0.60|G\rangle-0.13|S\rangle-0.79|N\rangle.
\label{eq:k}
\end{equation}

It is interesting and non-trivial that the pattern of decays determines
flavour mixing angles such that a  state having an ``octet" tendency
is sandwiched between two states that have a ``singlet" tendency.
As noted above and
elsewhere \cite{FC00} this is a potential signal for $G$ mixing
with a $q \overline q$ nonet. The output masses for $M_N$ and $M_S$
are consistent with the $K^*(1430)$ being in the nonet
and with the glueball
mass being at the lower end of the quenched lattice range
(see also \cite{bali,michael}). The mixing strength
also is in accord with lattice estimates\cite{Wein}.

The elements in eqs.(\ref{eq:i}-\ref{eq:k}) form the matrix $U$ as
defined in eqs.(\ref{eq:e},\ref{eq:f}). We have calculated the error on each
element taking into account the
correlated errors on their constituents which gives
\begin{equation}
\Delta U=\left(\begin{array}{ccc}
0.14 & 0.12 & 0.08\\
0.07 & 0.06 & 0.08\\
0.08 & 0.04 & 0.09
\end{array}\right).
\label{eq:l}
\end{equation}
This shows that the ``singlet-octet-singlet" phase pattern is robust.
The most sensitive probe of flavours and phases is in $\gamma \gamma$
couplings. In the spirit of ref.~\cite{re:CFL}, ignoring mass-dependent
effects, the above imply

\begin{eqnarray}
\Gamma(f_1(1710)\rightarrow \gamma\gamma):\Gamma(f_1(1500)\rightarrow
\gamma\gamma):\Gamma(f_1(1370)\rightarrow \gamma\gamma)=\nonumber\\
(5z_1+\sqrt{2}y_1)^2:(5z_2+\sqrt{2}y_2)^2:(5z_3+\sqrt{2}y_3)^2
=3.8:6.8:16.6.
\label{eq:m}
\end{eqnarray}
\par
The $\gamma \gamma$
width of $f_0(1500)$ exceeding that of $f_0(1710)$ arises
because the glueball is nearer to the $N$ than
the $S$. Contrast previous works where the $G$ was nearer to (or
even above) the
$S$, in which case the $f_0(1500)$ has the smallest $\gamma \gamma$
coupling of the three states \cite{re:CFL}.
 This shows how these $\gamma \gamma$
couplings have the potential to pin down the input pattern.

An interesting feature is the small value of the
pseudoscalar mixing angle ($\phi$); it is interesting that this
value agrees with recent work that has $\phi(\eta) \neq \phi(\eta')$
 \cite{FRERE}). We have checked that our results are not sensitive to allowing
these angles to be independent. If instead we
fix the value of $\phi$ to
-19$^\circ$ degrees, the $\chi^2$ of the fit
increases from 3.0 to 7.7 and the results are given in
tables~\ref{ta:BR} and \ref{ta:res1} respectively.
As can be seen the parameters of the fit are not very much
affected.
\par
We have also tried setting $r_3$ to zero:
the $\chi^2$ of the fit
increases to 13.9 (see
tables~\ref{ta:BR} and \ref{ta:res1}) and once again the parameters
are not much altered.
\par

\par
Other authors have claimed that
$M_G>M_S>M_N$~\cite{Wein, Stro, Bura}.
This scenario is disfavoured as,
if in the fit we require
$M_G>M_S>M_N$, the $\chi^2$ increases to 57.
In any event,
we are cautious about such claims~\cite{Wein, Stro, Bura}
as they are likely to
be significantly distorted by the presence of a higher, nearby,
excited $n \bar{n}$ state ($N^*$) such that $M_{N^*}>M_G>M_S$:
the philosophy of
dominant mixing with the nearest neighbours would then lead
again to the ``singlet - octet - singlet"
scenario that we have found above.
We defer detailed
discussion to a more complete report.
\par

Our preferred  solution has two further implications for the production
of these states
in $p \bar{p}$ annihilations,
in central $pp$ collisions and in radiative $J/\psi$ decays
that are in accord with data. These are interesting in that they
are consequences of the output and were not used as constraints.

The production of the $f_0$ states in $p \overline p \to \pi + f_0$ is expected
to be dominantly through the $n \overline n$ components of the
$f_0$ state, possibly through $gg$, but not prominently through
the $s \overline s$ components. (The possible presence of hidden
$s \overline s$ at threshold, noted by \cite{ellis} is in general
swamped by the above, and in any event appears unimportant in flight).
The above mixing pattern implies that

\begin{equation}
\sigma(p \overline p \to \pi + f_0(1710)) <
\sigma(p \overline p \to \pi +  f_0(1370)) \sim
\sigma(p \overline p \to \pi +  f_0(1500))
\end{equation}
Experimentally~\cite{thoma} the relative production rates are,

\begin{equation}
p \overline p \to \pi + f_0(1370) : \pi + f_0(1500)) \sim
1 : 1.
\end{equation}
and there is no evidence for the production of the $f_0(1710)$.
This would be natural if the production were via the
$n \overline n$ component.

For central production,
the cross sections of well established
quarkonia in WA102 suggest that the
production of $s \overline s$ is strongly suppressed \cite{pipikkpap}
relative to $n \overline n$.
The relative
cross sections for
the three states of interest here are

\begin{equation}
p p \to pp + ( f_0(1710):
 f_0(1500) :f_0(1370)) \sim 0.14:1.7:1.
\end{equation}
This would be natural if the production were via the
$n \overline n$ and $gg$ components.
\par
In addition,
the WA102 collaboration has studied the
production of these states as a function of
the azimuthal angle $\phi$, which is defined as the angle between the $p_T$
vectors of the two outgoing protons.
An important qualitative characteristic of these data is that
the $f_0(1710)$ and $f_0(1500)$ peak as
$\phi \to 0$ whereas the $f_0(1370)$ is more peaked
as $\phi \to 180$~\cite{WAphi}.
If the $gg$ and $n \overline n$
components are produced coherently as $\phi \to 0$ but out of phase
as $\phi \to 180$, then  this pattern of $\phi$ dependence and relative
production rates would follow; however, the relative coherence of
$gg$ and $n \overline n$ begs a dynamical explanation.

\par
In $J/\psi$ radiative decays, the absolute rates depend
sensitively on the phases and relative strengths of the
$G$ relative to the $q \overline q$ component, as well as the
relative phase of $n\bar{n}$ and $s\bar{s}$ within the latter.
The general pattern though is clear. Following the discussion
in ref.~\cite{re:CFL} we expect that the coupling to $G$ will be large;
coupling to $q\bar{q}$ with ``octet tendency" will be suppressed; coupling
to $q\bar{q}$ with ``singlet tendency" will be intermediate. Hence the
rate for $f_0(1370)$ will be smallest as the $G$ interferes destructively
against the $q\bar{q}$ with ``singlet tendency". Conversely, the
$f_0(1710)$ is enhanced by their constructive interference. The $f_0(1500)$
contains $q\bar{q}$ with ``octet tendency" and its production will be driven
dominantly by its $G$ content. If the $G$ mass is nearer to the $N$ than
to the $S$, as our results suggest, the $G$ component in $f_0(1500)$ is
large and cause the $J/\psi \to \gamma f_0(1500)$ rate to be comparable to
$J/\psi \to \gamma f_0(1710)$.
\par
In ref.~\cite{dunwoodie}, the branching ratio of
BR$(J/\psi \rightarrow \gamma f_0)(f_0\rightarrow \pi\pi + K \bar{K})$
for the $f_0(1500)$ and $f_0(1710)$ is presented. Using the WA102
measured
branching fractions~\cite{etaetapap}
for these resonances and assuming that all
major decay modes have been observed, the total relative production
rates in radiative $J/\psi$ decays can be calculated to be:
\begin{equation}
J/\psi \rightarrow f_0(1500) : J/\psi \rightarrow f_0(1710)
= 1.0 : 1.1 \pm 0.4
\end{equation}
which is consistent with the prediction above based on
our mixed state solution.
\par
In these mixed state solutions,
both
the $f_0(1500)$ and $f_0(1710)$ have
$n \bar{n}$ and $s \bar{s}$ contributions and so
it would be expected that both would be produced in
$\pi^-p$ and $K^-p$ interactions.
The $f_0(1500)$
has clearly been observed in
$\pi^-p$ interactions: it was first observed in the
$\eta\eta$ final state, although at that time it was referred to as
the $G(1590)$~\cite{NA12ETAETA}.
There is also evidence for the production
of the $f_0(1500)$ in $K^-p \rightarrow K^0_SK^0_S \Lambda$~\cite{8GEV,LASS}.
The signal is much weaker compared to the
well known $s \bar{s}$ state the $f_2^\prime(1525)$, as expected with our
mixings in eq.(\ref{eq:j}) and the suppressed $K\bar{K}$ decay associated
with the
destructive $n\bar{n} - s\bar{s}$ phase in the wavefunction.
\par
There is evidence for the $f_0(1710)$ in the reaction
$\pi^-p \rightarrow K^0_SK^0_Sn$,
originally called the $S^{*\prime}(1720)$~\cite{ETKIN,BOLONKIN}.
One of the longstanding problems of the $f_0(1710)$ is that
inspite of its dominant $K \bar{K}$ decay mode it was
not observed in $K^-p$ experiments~\cite{LASS,GAV}.
However, these concerns were based on the fact that initially the
$f_0(1710)$ had $J$~=~2.
In fact,
in ref.~\cite{lindenbaum} it was demonstrated that
if the $f_0(1710)$ had $J$~=~0, as it has now been found to have,
then the contribution in $\pi^-p$ and $K^-p$ are compatible.
One word of caution should be given here: the analysis
in ref.~\cite{lindenbaum} was performed with a $f_0(1400)$ not
a $f_0(1500)$ as we today know to be the case. As a further test of
our solution, it would
be nice to see the analysis of ref.~\cite{lindenbaum}
repeated with the mass and width of the
$f_0(1500)$ and the decay parameters of the $f_0(1710)$
determined by the WA102 experiment.
\par
In summary,
based on the hypothesis that
the scalar glueball
mixes with the nearby $q \overline q$ nonet states,
we have determined the flavour content of the
$f_0(1370), f_0(1500)$ and $f_0(1710)$
by studying their decays into all pseudoscalar meson pairs.
The solution we have found
is also compatible with the relative production
strengths of the $f_0(1370), f_0(1500)$ and $f_0(1710)$
in $pp$ central production, $p \bar{p}$ annihilations
and $J/\psi$ radiative decays.

\begin{center}
{\bf Acknowledgements}
\end{center}
\par
This work is supported, in part, by grants from
the British Particle Physics and Astronomy Research Council,
the British Royal Society,
and the European Community Human Mobility Program Eurodafne,
contract NCT98-0169.

\newpage
\begin{table}[h]
\caption{The measured and predicted branching ratios
with the individual $\chi^2$ contributions coming from the fits.}
\label{ta:BR}
\vspace{0.2in}
\begin{center}
\begin{tabular}{|c|c|cc|cc|cc|} \hline
 & & & &&&&\\
 & Measured & \multicolumn{2}{c|}{All free}
 & \multicolumn{2}{c|}{$\phi$~=~-19$^\circ$}
 & \multicolumn{2}{c|}{$r_3$= 0} \\
 & Branching &Fitted& $\chi^2$ &Fitted& $\chi^2$ &Fitted& $\chi^2$\\
 & ratio& & &&&&\\
 & & & &&&&\\ \hline
  & & & &&&&\\
 $\frac{f_0(1370)\rightarrow \pi \pi}{f_0(1370)\rightarrow K \overline K}$
&2.17 $\pm$ 0.9 & 2.13 &0.001 & 2.1 & 0.004 &2.25 & 0.007 \\
 & & & &&&&\\
 $\frac{f_0(1370)\rightarrow \eta \eta}{f_0(1370)\rightarrow K \overline K}$
&0.35 $\pm$ 0.21 & 0.41 & 0.1 & 0.01 & 2.6 & 0.01 & 2.6 \\
 & & & &&&&\\
 $\frac{f_0(1500)\rightarrow \pi \pi}{f_0(1500)\rightarrow \eta \eta}$
&5.5 $\pm$ 0.84 & 5.60 & 0.01 &5.6 & 0.02 &6.20 & 0.69\\
 & & & &&&&\\
 $\frac{f_0(1500)\rightarrow K \overline K}{f_0(1500)\rightarrow \pi \pi}$
&0.32 $\pm$ 0.07 & 0.37 & 0.54 &0.32& 0.001 & 0.35 & 0.22\\
 & & & &&&&\\
 $\frac{f_0(1500)\rightarrow \eta \eta^\prime}{f_0(1500)\rightarrow \eta
\eta}$
&0.52 $\pm$ 0.16 & 0.60 & 0.23 &0.5 & 0.01 &0.20 & 3.9\\
 & & & &&&&\\
 $\frac{f_0(1710)\rightarrow \pi \pi}{f_0(1710)\rightarrow K \overline K}$
&0.20 $\pm$ 0.03 & 0.19 & 0.05  &0.20 & 0.002 &0.19 & 0.08\\
 & & & &&&&\\
 $\frac{f_0(1710)\rightarrow \eta \eta}{f_0(1710)\rightarrow K \overline K}$
&0.48 $\pm$ 0.14 & 0.29 & 1.9 & 0.17 & 4.9 & 0.13 & 6.1 \\
 & & & &&&&\\
 $\frac{f_0(1710)\rightarrow \eta \eta^\prime}{f_0(1710)\rightarrow \eta
\eta}$
&$< 0.05 (90 \% cl)$& 0.034  & 0.27 & 0.04 & 0.05 & 0.06 & 0.05\\
 & & & &&&&\\ \hline
\end{tabular}
\end{center}
\end{table}
\begin{table}[h]
\caption{The ratio of decay amplitudes squared.}
\label{ta:amp}
\vspace{0.2in}
\begin{center}
\begin{tabular}{|c|c|c|} \hline
 & &  \\
 & $\eta \eta/\pi\pi$ &$\eta\eta/K \overline K$\\
 & & \\ \hline
  & & \\
 $f_0(1370)$ &0.74 $\pm$ 0.51 & 1.64 $\pm$ 0.96 \\
 & & \\
 $f_0(1500)$ &0.68 $\pm$ 0.11 & 2.42 $\pm$ 0.64 \\
 & & \\
 $f_0(1710)$ &7.9 $\pm$ 2.4 & 1.96 $\pm$ 0.64 \\
  & & \\ \hline
\end{tabular}
\end{center}
\end{table}
\begin{table}[h]
\caption{The theoretical reduced partial widths.}
\label{ta:RPW}
\vspace{0.2in}
\begin{center}
\begin{tabular}{|c|c|} \hline
 &  \\
$\gamma^2(f_i\rightarrow \eta\eta^\prime)$
&$2[2\alpha\beta(z_i - \sqrt{2}y_i) + 2\alpha \beta x_i r_3]^2$ \\
 &  \\
$\gamma^2(f_i\rightarrow \eta\eta)$
&$[2\alpha^2z_i+2\sqrt{2}\beta^2y_i+r_2x_i+2\beta^2 x_i r_3]^2$ \\
 &  \\
$\gamma^2(f_i \rightarrow \pi\pi)$
&$3[z_i+r_2x_i]^2$ \\
 &  \\
$\gamma^2(f_i\rightarrow K\bar{K})$
&$4[\frac{1}{2}(z_i +\sqrt{2}y_i)+r_2x_i]^2$ \\
 & \\ \hline
\end{tabular}
\end{center}
\end{table}
\clearpage
\begin{table}[h]
\caption{The solutions
for the minimum $\chi^2$.}
\label{ta:res1}
\vspace{0.2in}
\begin{center}
\begin{tabular}{|c|c|c|c|}\hline
& & & \\
Parameters& All Free & $\phi$~=~ -19$^\circ$ & $r_3$~=~0 \\
& & & \\ \hline
& & & \\
$\chi^2$  &3.0 & 7.7 & 13.9 \\
& & & \\
$M_G$ (MeV) &1440 $\pm$ 16 &1433 $\pm$ 19 &1437 $\pm$ 15 \\
$M_S$ (MeV)&1672 $\pm$ 9&1668 $\pm$ 8&1672 $\pm$ 13\\
$M_N$ (MeV)&1354 $\pm$ 28&1366 $\pm$ 25&1368 $\pm$ 29\\
$M_3$ (MeV)&1256 $\pm$ 31&1251 $\pm$ 18&1264 $\pm$ 14\\
$f$ (MeV)&91 $\pm$ 11&95 $\pm$ 13&90 $\pm$ 11\\
$\phi$ (Deg)&-5 $\pm$ 4&-19&-25 $\pm$ 4\\
$r_2$ & 1.02 $\pm$ 0.14&0.92 $\pm$ 0.18&0.95 $\pm$ 0.12\\
$r_3$ & 1.04 $\pm$ 0.24&0.77$\pm$ 0.26&0  \\
& & & \\ \hline
\end{tabular}
\end{center}
\end{table}
\clearpage
{ \large \bf Figures \rm}
\begin{figure}[h]
\caption{
Possible Decays to Pseudoscalar meson pairs ($PP$).
a) The direct coupling of the $q \bar{q}$
to the $PP$ pair, b) the coupling of the $q \overline q$ to $PP$
via intermediate gluons, c) the coupling of the glueball
component to $PP$ and d) the direct coupling of gluons to
isoscalar mesons.
}
\label{fi:1}
\end{figure}
\begin{figure}[h]
\caption{
The Decays to Pseudoscalar meson pairs ($PP$) considered in this analysis.
a) The coupling of the $q \bar{q}$
to the $PP$ pair,
b) the coupling of the glueball
component to $PP$ and c) the direct coupling of gluons to
the gluonic component of the final state mesons.
}
\label{fi:2}
\end{figure}
\newpage
\begin{center}
\epsfig{figure=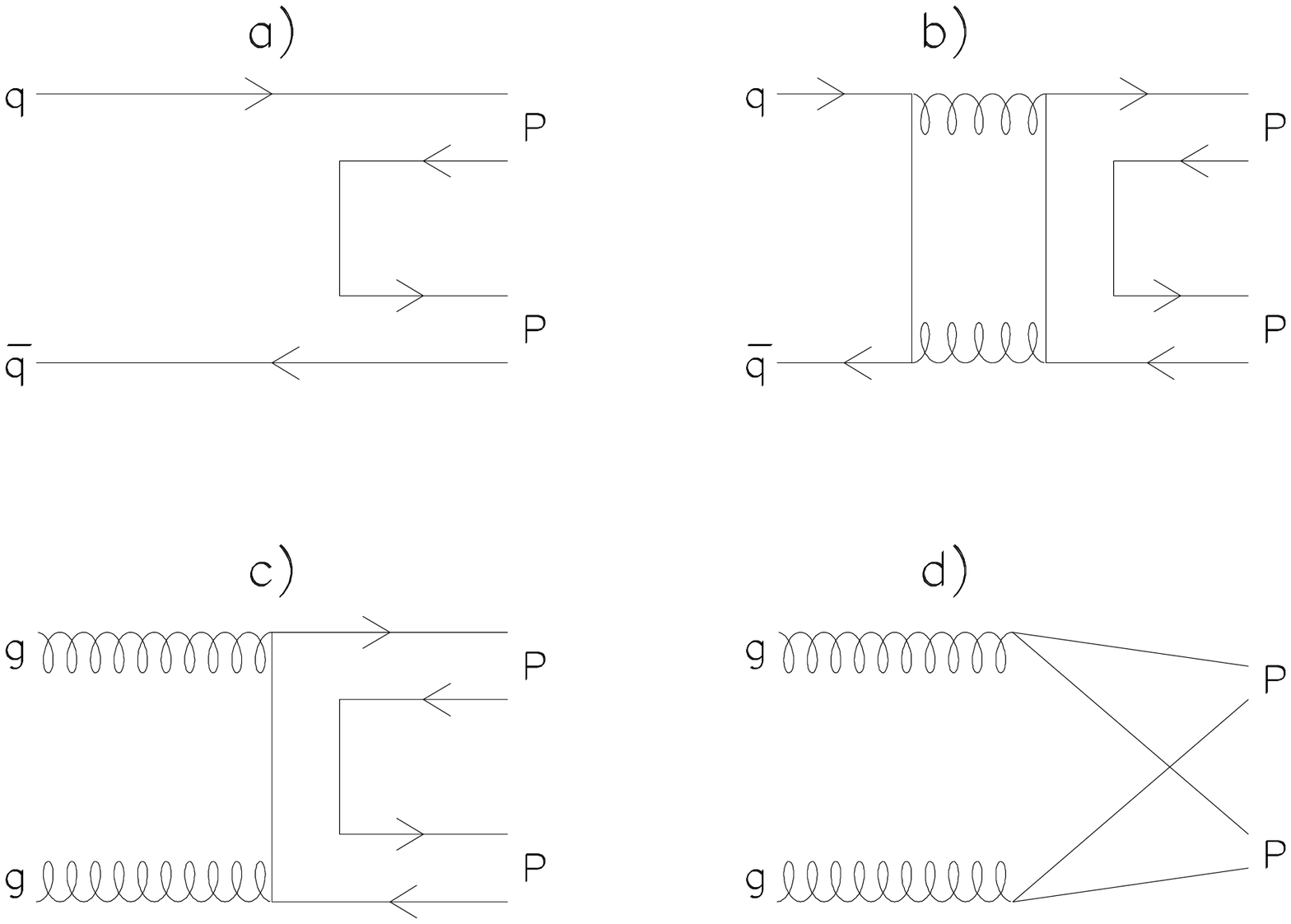,height=22cm,width=17cm}
\end{center}
\begin{center} {Figure 1} \end{center}
\newpage
\begin{center}
\epsfig{figure=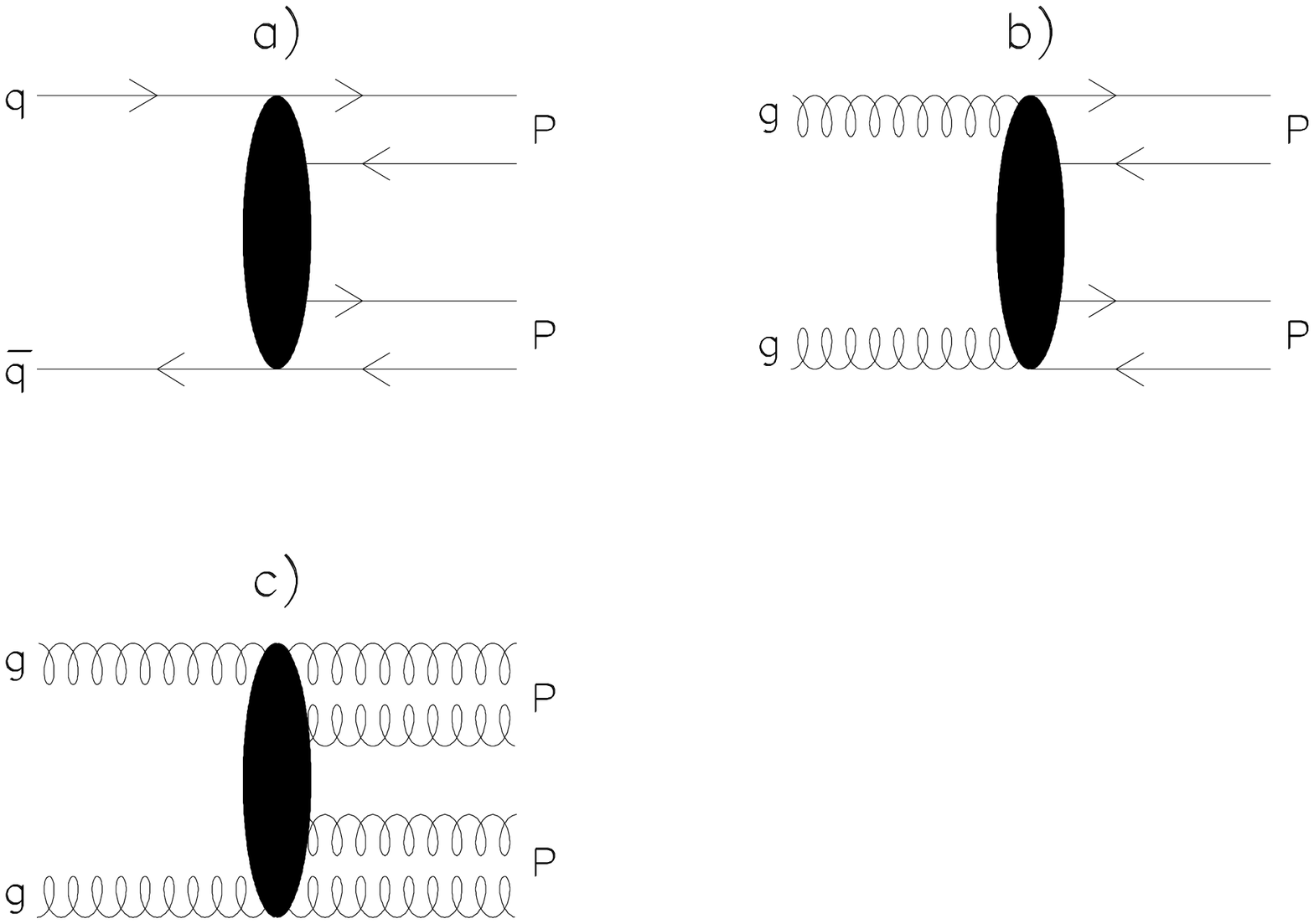,height=22cm,width=17cm}
\end{center}
\begin{center} {Figure 2} \end{center}
\newpage

\newpage
\begin{thebibliography}{99}
\bibitem{re:lgt}
G. Bali  {\em et al.} (UKQCD),
Phys. Lett.  {\bf B309}  (1993) 378;

D. Weingarten, hep-lat/9608070;

J. Sexton et al., Phys. Rev. Lett. {\bf 75} (1995) 4563;

F.E. Close and M.J. Teper, ``On the lightest Scalar Glueball"
Rutherford Appleton Laboratory report no. RAL-96-040;
Oxford University report no. OUTP-96-35P

W. Lee and D. Weingarten, hep-lat/9805029

G. Bali hep-ph/0001312


\bibitem{re:AC}

C. Amsler and F.E. Close Phys. Lett. {\bf B353 } \rm (1995) 385.
\bibitem{re:IP}
N. Isgur and J. Paton, Phys. Rev. {\bf D31} (1985) 2910.
\bibitem{anisovich}
V.V. Anisovich, Physics-Uspekhi {\bf 41} (1998) 419.
\bibitem{re:FC89}
F.E. Close, Rep. Prog .Phys. {\bf 51} (1988) 833.
\bibitem{re:CFL}
F.E. Close, G. Farrar and Z.P. Li, Phys.Rev. {\bf D55} (1997) 5749.


\bibitem{Wein}
D. Weingarten, Nucl. Phys. Proc. Suppl. {\bf 53} (1997) 232;
{\bf 63} (1998) 194;{\bf 73} (1999) 249.
\bibitem{mix}
De-Min Li, Hong Yu and Qi-Xing Shen, hep-ph/0001107.
\bibitem{Genov}
M. Genovese, Phys.Rev {\bf D46} (1992) 5204
\bibitem{FC00}
F.E. Close, Proc of MESON2000 (in preparation; unpublished)
%
\bibitem{etaetapap}
D. Barberis {\em et al.,} hep-ex/0003033 To be published in Phys. Lett. B.
\bibitem{gershtein}
S.S. Gershtein {\em et al.,} Zeit. Phys. {\bf C24} (1984) 305.
\bibitem{nodes}
T. Barnes, F.E. Close, P. Page and E. Swanson, Phys. Rev {\bf D55} (1997) 4157
\bibitem{Jaf00}
M. Alford and R.L. Jaffe, hep-lat/0001023
\bibitem{tornpenn}
N.A. Tornqvist, Zeit. Phys. {\bf C68} (1995) 647; \\
M. Boglione and M.R. Pennington, Phys. Rev. Lett. {\bf 79} (1997) 1998.
\bibitem{bali}
G. Bali  {\em et al.} (SESAM),
hep-lat/0003012.
\bibitem{michael}
C. Michael, M. Foster and C. McNeile, hep-lat/9909036
\bibitem{FRERE}
R. Escribano and J.-M. Frere, Phys. Lett. {\bf B459} (1999) 288
%
\bibitem{Stro}
M. Strohmeier-Presicek, T. Gutsche, R. Vinh
Mau, Amand Faessler, Phys. Rev. {\bf D60} (1999) 054010.
\bibitem{Bura}
L. Burakovsky, P.R. Page, Phys. Rev.{\bf D59} (1999) 014022.
\bibitem{ellis}
J. Ellis, E. Gabathuler and M. Karliner, Phys. Lett. {\bf B217} (1989) 173.
\bibitem{thoma}
U. Thoma, Proceedings of Hadron 99, Beijing, China 1999.
\bibitem{pipikkpap}
D. Barberis {\em et al.,} Phys. Lett. {\bf B462} (1999) 462.
\bibitem{WAphi}
D. Barberis {\em et al.,} Phys. Lett. {\bf B467} (1999) 165.
%
\bibitem{dunwoodie}
W. Dunwoodie, Proceedings of Hadron 97, AIP Conf. Series 432 (1997) 753.
%
\bibitem{NA12ETAETA}
F. Binon {\em et al.,} Il Nuovo Cimento {\bf A78} (1983) 313.
\bibitem{8GEV}
M. Baubillier {\em et al.,} Zeit. Phys. {\bf C17} (1983) 309.
\bibitem{LASS}
D. Aston {\em et al.,} Nucl. Phys. {\bf  B301} (1988) 525.
\bibitem{ETKIN}
A. Etkin {\em et al.,} Phys. Rev. {\bf D25} (1982) 1786.
\bibitem{BOLONKIN}
B.V. Bolonkin {\em et al.,} AIP. Conf. Proc. 185 (1988) 289.
\bibitem{GAV}
Ph. Gavillet {\em et al.,} Zeit. Phys. {\bf C16} (1982) 119.
\bibitem{lindenbaum}
S. Lindenbaum and R.S. Longacre, Phys. Lett. {\bf B274} (1992) 492.
\end{thebibliography}
\end{document}